\documentclass[aps,prb,nobibnotes,twocolumn,superscriptaddress,shortbibliography]{revtex4-2}

\usepackage{amsfonts}
\usepackage{mathrsfs}
\usepackage{amsmath}% needed for subequations
\usepackage{color}
\usepackage{graphicx}
\usepackage{bm}% bold maths
\usepackage{amssymb}
\usepackage{xspace}
\usepackage{epstopdf}
\usepackage{dcolumn}% Align table columns on decimal point
\usepackage{longtable}
\usepackage{multirow}
\usepackage{float}
\usepackage{comment}
\usepackage{lineno}
\usepackage[colorlinks=true, letterpaper=true, pdfstartview=FitV,  linkcolor=blue, citecolor=blue, urlcolor=blue]{hyperref}

\makeatother

\begin{document}

\title{Photoinduced topological phase transition in monolayer Ti$_2$SiCO$_2$}

\author{Pu Liu}
\email{liupu@bit.edu.cn}
\affiliation{Centre for Quantum Physics, Key Laboratory of Advanced Optoelectronic Quantum Architecture and Measurement (MOE), School of Physics, Beijing Institute of Technology, Beijing, 100081, China}
\affiliation{Beijing Key Lab of Nanophotonics \& Ultrafine Optoelectronic Systems, School of Physics, Beijing Institute of Technology, Beijing, 100081, China}

\author{Chaoxi Cui}
\affiliation{Centre for Quantum Physics, Key Laboratory of Advanced Optoelectronic Quantum Architecture and Measurement (MOE), School of Physics, Beijing Institute of Technology, Beijing, 100081, China}
\affiliation{Beijing Key Lab of Nanophotonics \& Ultrafine Optoelectronic Systems, School of Physics, Beijing Institute of Technology, Beijing, 100081, China}

\author{Zhi-Ming Yu}
%\email{zhiming\_yu@bit.edu.cn}
\affiliation{Centre for Quantum Physics, Key Laboratory of Advanced Optoelectronic Quantum Architecture and Measurement (MOE), School of Physics, Beijing Institute of Technology, Beijing, 100081, China}
\affiliation{Beijing Key Lab of Nanophotonics \& Ultrafine Optoelectronic Systems, School of Physics, Beijing Institute of Technology, Beijing, 100081, China}

\begin{abstract}
The TiSiCO-family monolayer $X_2Y$CO$_2$($X$=Ti, Zr, Hf; $Y$=Si, Ge) is a two-dimensional second-order topological insulator with unique valley-layer coupling in equilibrium condition.
In this work, based on the four-band tight-binding (TB) model of monolayer Ti$_2$SiCO$_2$ (ML-TiSiCO) and the Floquet theory, we study the non-equilibrium properties of the ML-TiSiCO under a periodic field of laser and a gate-electric field.
We find the interaction between the time-periodic polarized light and the electric field can lead to a variety of intriguing topological phase transitions.
By driving the system with only circularly polarized light (CPL), a photoinduced topological phase transition occurs from a second-order topological insulator to a Chern insulator with a Chern number of $C=\pm$2, and the sign of the Chern number $C$ is determined by the chirality of the incident light.
%This indicates that the band structure is modified by photon dressing with a new dispersion, revealing that the topological properties of the ML-TiSiCO can be altered by varying parameters such as the amplitude and direction of the incident light. 
Further adding a perpendicular electric field, we find that the ML-TiSiCO exhibits a rich phase diagram, consisting of Chern insulators with different Chern numbers and various topological semimetals. 
In contrast, since the linearly polarized light (LPL) does not break time-reversal symmetry, the Chern number of the system would not be changed under the irradiation of LPL.
However, there still exist many topological phases, including second-order topological insulator, topological semi-Dirac, Dirac and valley-polarized Dirac semimetals under the interaction between the  LPL and the  electric field.
Our results not only enhance the understanding of the fundamental properties of ML-TiSiCO  but also broaden the potential applications of such materials in optoelectronic devices.
\end{abstract}

\maketitle
\section{Introduction}

Valleytronics is an emerging field that focuses on the manipulation of the valley degree of freedom in materials. The valleytronics materials are characterized by the presence of multiple energy extremes in the low-energy region that are connected by symmetry, giving rise to additional valley degrees of freedom~\cite{JohnR,Hongyi,StevenA,Rycerz,Gunawan,XiaoDi,YaoWang,XiaoDi2,ZengweiZhu,CaiTianyi}.
The flexibility and controllability of two-dimensional (2D) valleytronics materials have promoted the development of basic research and device applications in the field of valleytronics~\cite{GuiBinLiu,JohnR,StevenA,PanHui,LiuDaPing, PanHui2,WenYiTong,SettnesMikke,Hongyi}. In addition to traditional graphene and transition metal chalcogenides (TMDs), more 2D valleytronics materials with novel physical properties have emerged recently, such as 2H-VSe$_2$, Nb$_3$I$_8$, 2H-FeCl$_2$, VSi$_2$N$_4$, Cr$_2$Se$_3$, TiBr$_2$, GeSe, and so on~\cite{WenYi,PengRui,Qirui,Zhonglin,JiangPeng,ChengHaiXia,ZhaoYiFeng,ZhaoXiuwen,XinWeiShen}.

Among all these valleytronics materials, the TiSiCO-family monolayers $X_2Y$CO$_2$ ($X$=Ti, Zr, Hf; $Y$= Si, Ge) is outstanding due to its unique valley-layer coupling (VLC), which allows direct and efficient control of the valley by gate-electric field~\cite{ZMYuValleytronics}.
The monolayer Ti$_2$SiCO$_2$ (ML-TiSiCO) is characterized by two valleys located at high-symmetry points ($X$ and $Y$) in the Brillouin zone (BZ)~\cite{ZMYuValleytronics}. Unlike graphene and TMDs, the two valleys are located at time-reversal ($\cal{T}$) invariant points and are connected by the spatial operators $S_{4z}$ and $C_{2,110}$, rather than $\cal{T}$.
One unique feature of the ML-TiSiCO is that the conduction (valence) bands of the system in different valleys possess strong but opposite layer polarization, leading to a special kind of VLC. With this VLC, switchable valley polarization can be realized by applying a gate-electric field, which is highly desirable for applications.
A dynamical generation of valley polarization also can be achieved in the ML-TiSiCO, due to the valley-contrasting linear dichroism~\cite{ZMYuValleytronics}.
Moreover, it has been predicted that the ML-TiSiCO is a
second-order topological insulator rather than a trivial semiconductor~\cite{YilinHan}. Therefore, under the control of optical and electrical fields, we can anticipate the emergence of more novel properties in the ML-TiSiCO.

%The unique dispersion relationships in the low-energy range of ML-TiSiCO can lead to novel optical characteristics and responses.
Optical pumping in quantum materials has emerged as a promising approach for exploring novel properties of matter that are not present in equilibrium systems~\cite{MarkS,Takashi2,ChanghuaBao}. For instance, the Floquet band engineering uses periodically driven external fields to manipulate electric states, and even to create novel topological states~\cite{FuLiang,OkaTakashi,EdbertJSie,YHWANG,WangRui,WangRui2,Donghuixu,Donghuixu2,Donghuixu3,Donghuixu4}. 
Numerous theoretical studies have proposed that employing periodic monochromatic pumping on ultrafast timescales can lead to non-equilibrium photon-modified phases with topologically protected edge states~\cite{Netanel,Luca,OkaTakashi,FuLiang,ZhuTongshuai,YangQi,Seshadri,AHuaman}.
The irradiation of off-resonant circularly polarized light (CPL) induces the opening of the band gap in graphene, leading to the emergence of chiral edge states~\cite{FuLiang,OkaTakashi}. By irradiation of CPL at fixed electric field, silicene undergoes a topological transition from a topological insulator to a photoinduced quantum Hall insulator or spin-polarized quantum Hall insulator~\cite{Ezawa}. Similarly, the TMD heterobilayers driven by CPL, can exhibit the valley quantum spin Hall effect which has usually been limited to magnetic systems ~\cite{ZhanFangyang}. Various nonequilibrium topological phases have been discovered in black phosphorus under irradiation of polarized light~\cite{LiuHang,Taghizadeh,KangYousung}. Elliptic polarized fields are likely to drive the black phosphorus into a quantum anomalous Hall phase, while a semimetallic phase will be induced by linearly polarized light (LPL)~\cite{Dutreix}. In particular, the nonequilibrium quantum transport properties can be controlled by the periodic light field, offering the potential to optically control nonequilibrium quantum properties~\cite{FuLiang,OkaTakashi}. The topological transport properties can be induced in graphene driven by CPL, resulting in the quantum Hall effects in this system~\cite{FuLiang}. In strained black phosphorus excited by the incident light, the topological surface states also exhibit nonequilibrium transport properties which is governed by the chirality of CPL~\cite{LiuHang}. Further, some photoinduced topological states and phase transitions have been confirmed in experiments~\cite{EdbertJSie,YHWANG,Edbert}.

In this work, based on a four-band tight-binding (TB) model and the Floquet theory, we theoretically study the topological phase transition of the ML-TiSiCO driven by the periodic external field and electrostatic gate field. We find that under the optical and electric fields, various topological phases can be achieved in ML-TiSiCO, which offers an appealing material platform to explore the photoinduced topological states and their optical-electric properties. Under the irradiation of CPL with increasing intensity, the ML-TiSiCO undergoes a transition from a second-order topological insulator to a Chern insulator, accompanied by a change in the Chern number from $C=0$ to $C=\pm 2$. We also have obtained two chiral edge states in the band structure of the nanoribbon extended in [100] direction.
The chirality of the edge states, as well as the sign of the Chern number of the driven system, can be switched by the chirality of CPL. Further, under CPL and perpendicular electric field $E_z$, we have obtained a rich topological phase diagram, which includes Chern insulators with different Chern numbers $C$ and multiple kinds of topological semimetals.
We also investigate the effect of the interaction between LPL and the electric field on the ML-TiSiCO. Since LPL keeps the time-reversal symmetry, the Chern number of the system always remains $C=0$ under its irradiation.
However, the system still undergoes many topological phase transitions, transforming into multiple kinds of topological semimetals, including topological semi-Dirac, Dirac, and valley polarized Dirac semimetals. 
Furthermore, the types and positions of these crossing points in the BZ can be altered by controlling the direction and intensity of LPL and electric field.

This paper is organized as follows. In Sec. \ref{sec II}, based on the four-band TB model and Floquet theory, we obtain the effective Flouqet  Hamiltonian driven by the periodic off-resonant light field. In Sec. \ref{sec:III}, we study the topological phase transitions of the ML-TiSiCO under the irradiation of CPL and LPL, as well as the presence of a gate electric field. We demonstrate rich phase diagrams by adjusting CPL and electric field. Conclusions are given in Sec. \ref{sec:IV}.

\section{The effective Floquet Hamiltonian of ML-TiSiCO}\label{sec II}

\begin{figure}
\includegraphics[width=1\columnwidth]{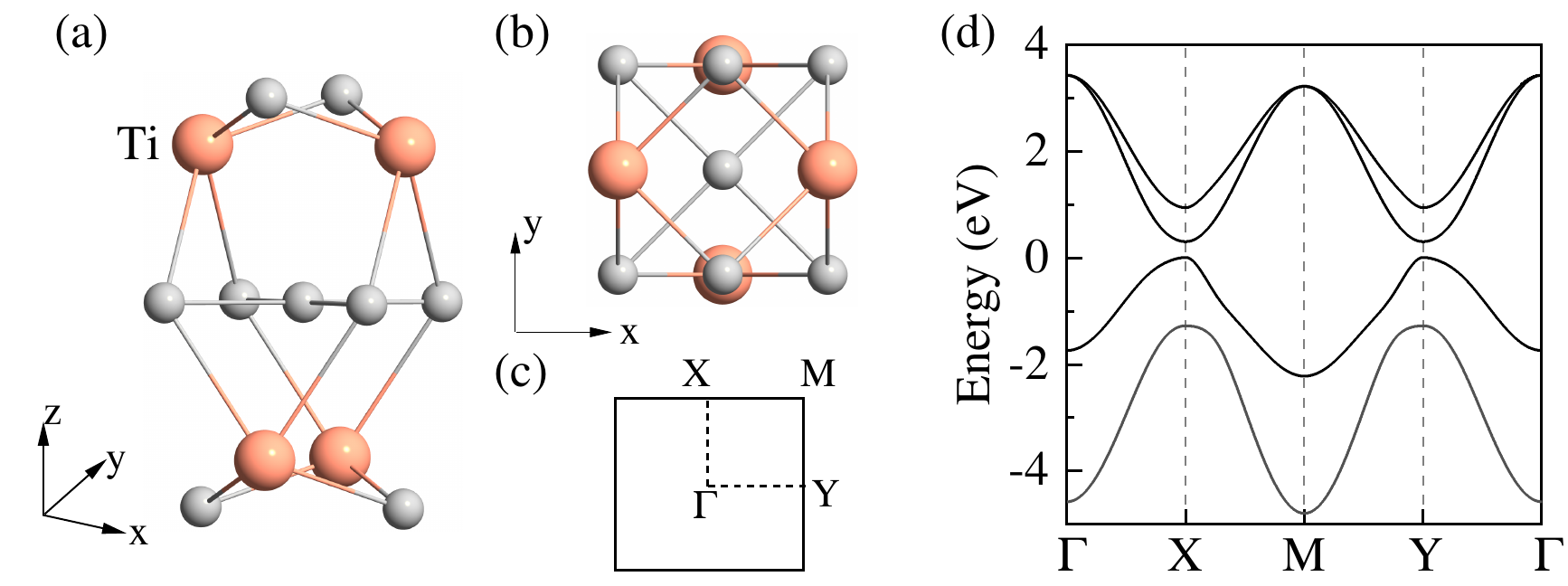}
\caption
{(a) Side view and (b) top view of the crystal structure for ML-TiSiCO. The orange atoms represent the Ti atoms, and the other atoms are represented by the gray atoms. (c) The corresponding Brillouin zone. (d) The band structure for undriven ML-TiSiCO. }
\label{Fig.1}
\end{figure}

As illustrated in Fig.~\ref{Fig.1}(a-b), the ML-TiSiCO possesses five square-lattice atomic layers vertically stacked in the sequence of O-Ti-Si/C-Ti-O. It belongs to layer group No. 59 (or space group No. 115) with $D_{2d}$ point group symmetry. Previous research has indicated that the low-energy states of ML-TiSiCO are primarily contributed by the $d$ orbitals $\{ d^t_{z^2}, d^t_{yz}, d^b_{z^2}, d^b_{xz}\}$  of the Ti atoms in the top and bottom layers.
Thus, the low-energy physics of the ML-TiSiCO can be effectively described by the following four-band TB model \cite{ChaoxiCui},
\begin{eqnarray}
H_{int}=\left(\begin{array}{cc}
H_{\mathrm{top}} & H_{\mathrm{inter}}\\
H_{\mathrm{inter}}^{\dagger} & H_{\mathrm{bottom}}
\end{array}\right)
\end{eqnarray}
where $H_{\mathrm{bottom}}(k_{x},k_{y})=H_{\mathrm{top}}(k_{y},-k_{x})$ with $H_{top}$
\begin{eqnarray*}
H_{top}=\Delta \sigma_z
+\left[
\begin{array}{cc}
\begin{array}{c}
t_{1x}(f_{1,x}+f_{1,x}^{\ast }) \\
+t_{1y}(f_{1,y}+f_{1,y}^{\ast })%
\end{array}
& t_{3}(f_{1,y}-f_{1,y}^{\ast }) \\
-t_{3}(f_{1,y}-f_{1,y}^{\ast }) &
\begin{array}{c}
t_{2x}(f_{1,x}+f_{1,x}^{\ast }) \\
+t_{2y}(f_{1,y}+f_{1,y}^{\ast })%
\end{array}%
\end{array}%
\right],
\end{eqnarray*}
and 
\begin{eqnarray}
H_{\mathrm{inter}}=\left[\begin{array}{cc}
t_{4}\left(f_{2}+f_{2}^{*}+f_{3}+f_{3}^{*}\right) & -t_{5}\left(f_{2}-f_{2}^{*}+f_{3}-f_{3}^{*}\right)\\
-t_{5}\left(f_{2}-f_{2}^{*}-f_{3}+f_{3}^{*}\right) & t_{6}\left(f_{2}+f_{2}^{*}-f_{3}-f_{3}^{*}\right)
\end{array}\right],
\end{eqnarray}
where $\sigma_z$ is the Pauli matrix, $f_{1,x}=e^{ik_x},f_{1,y}=e^{ik_y},f_{2}=e^{i\left(\frac{k_{x}}{2}+\frac{k_{y}}{2}\right)}$ and $f_{3}=e^{i\left(\frac{k_{x}}{2}-\frac{k_{y}}{2}\right)}$. $\Delta$ = 0.036 eV is the on-site energies of the two orbits of the Ti atoms. $t_{1x}$ = -0.729 eV, $t_{1y}$ = -0.866 eV, $t_{2x}$ = 0.621 eV, $t_{2y}$ = 1.113 eV, and $t_3$ = 0.677 eV denote the hoppings between the Ti atoms in the same layer, while $t_4$ = -0.355 eV, $t_5$ = 0.127 eV, and $t_6$ = 0.322 eV represent the hoppings between the Ti atoms in the top and bottom layers. The band structure of ML-TiSiCO  is shown in Fig.~\ref{Fig.1}(d) and the corresponding BZ is shown in Fig.~\ref{Fig.1}(c).

As shown in Fig.~\ref{Fig.1}(d), for the undriven ML-TiSiCO, there are two valleys at $X$ ($\pi$,0) and $Y$ (0,$\pi$) points in the BZ. Due to the strong VLC effect, applying a gate-electric field normal to the plane of the system is a convenient and efficient method to control the bands, as it can produce an opposite electrostatic potential for the top and bottom atoms \cite{ZMYuValleytronics}. The effect of the gate electric field can be incorporated into the TB model by introducing an on-site energy term,
\begin{eqnarray}
H_{E}=U\left[\begin{array}{cccc}
1 & 0 & 0& 0\\
0 & 1 & 0 &0\\
0 & 0 & -1 &0\\
0 & 0 & 0 &-1\\
\end{array}\right],
\label{Eq.3}
\end{eqnarray}
where $U$ represents the electrostatic potential energy applied to the Ti atoms, with opposite sign on the Ti atoms in the top and bottom layers. The sign of $U$ denotes the direction of the perpendicular electric field.

Here, we consider the off-resonant periodic light to drive the topological phase transition of ML-TiSiCO, which is also widely used in the Floquet engineering of other 2D materials~\cite{FuLiang,ChanghuaBao}. The off-resonant light or high-frequency light, of which the drive frequency $\omega$ is much larger than the bandwidth $W$, effectively modifies the electron band structures through virtual photo absorption processes~\cite{FuLiang,MarkS}. For concreteness, suppose that a light beam comes in the $z$ direction, with the vector potential $\mathbf{A}(t)=[\eta A_x \sin \omega t, A_y\cos(\omega t-\psi), 0]$ with $\mathbf{E}(t)=\partial \mathbf{A}(t)/\partial t$. This is a time-periodic light field with $\omega=\frac{2\pi}{T}$ and $\mathbf{A}(t+T)=\mathbf{A}(t)$. And $A_{x/y}$ is characterized by the dimensionless number $A_i=eE_i\omega^{-1}$ and adjustable in the experiment by simply tuning the light intensity $E_i$.
The $\psi=0$ is for the CPL and $\psi=\pi/2$ for the LPL, while the $\eta$=1 or -1 represents the right- or left-handed CPL. 
In practical calculations,
the periodic driving field is introduced in the Hamiltonian by using the minimal coupling substitution $\mathbf{k}\rightarrow \mathbf{k}-e\mathbf{A}(t)$.

In the limit where the driving frequency $\omega$ is much larger than the bandwidth $W$, according to the Magnus expansion, the effective Hamiltonian expanded up to $1/\omega$ reads~\cite{MarinBukov,Kitagawa1,Kitagawa2}
\begin{eqnarray}
H_{eff}=H_{0}+\Delta H=H_{0}+\frac{1}{\omega}[H_{-1},H_{1}],
\label{Eq.4}
\end{eqnarray}
where $H_n=\frac{1}{T}\int_0^T H_{int}(t)e^{in\omega t}dt$ is a time-averaged Hamiltonian, which represents laser-assisted processes, absorbing or emitting $n$ photons. In the high-frequency approximation, we find that the forms of $H_n$($n=0,\pm1$) are similar to $H_{int}$, with the distinction that their hoppings parameters have been modified. For the $H_0$, we have
\begin{eqnarray*}
&&f_{1,x}\rightarrow f_{1,x}^{0}=J_{0}(A_{x})f_{1,x},f_{1,y}\rightarrow f_{1,y}^{0}=J_{0}(A_{y})f_{1,y},\\
&&f_{2}\rightarrow f_{2}^{0}=J_{0}(G_{+})f_2,f_{3}\rightarrow f_{3}^{0}=J_{0}(G_{-})f_3\\
\end{eqnarray*}
and for the $H_{1}$, we have  
\begin{eqnarray*}
&&f_{1,x}\rightarrow f_{1,x}^{1}=\eta J_{1}(A_{x})f_{1,x},\\
&&f_{1,y}\rightarrow f_{1,y}^{1}=e^{i\left(\psi-\frac{\pi}{2}\right)}J_{1}(A_{y})f_{1,y},\\
&&f_{2}\rightarrow  f_{2}^{1}=e^{-i\varPsi_{+}}J_{1}(G_{+})f_2,\\
&&f_{3}\rightarrow f_{3}^{1}=e^{-i\varPsi_{-}}J_{1}(G_{-})f_3,
\end{eqnarray*}
where $J_m(A_i)(m=0,1)$ represents the 0-th or 1-st Bessel function and another parameters
$\tan\varPsi_{\pm}=\frac{A_{y}\cos\psi}{\pm A_{x}\eta+A_{y}\sin\psi}$ and $G_{\pm}=\eta\sqrt{\frac{A_{y}^{2}}{4}+\frac{A_{x}^{2}}{4}\pm\frac{1}{2}\eta A_{x}A_{y}\sin\psi}$. Furthermore, $H_{-1}$ and $H_{1}$ are conjugate.
The coupling between the incident light and the electrons can be considered as the additional phase factor in the hopping parameters. For the first term $H_0$, the effect of CPL essentially involves the renormalization of the original hopping parameters, with the hopping coefficients being modified by Bessel functions. Since Bessel functions usually yield values less than or equal to 1, this renormalization process typically results in a decrease or even complete elimination of the amplitudes of the original nearest-neighbor and next-nearest-neighbor hoppings. By calculating the commutation relations in $\Delta H$, it becomes clear that the second term in the effective Hamiltonian Eq.~(\ref{Eq.4}) introduces additional long-distance hoppings. Notice that these additional hopping parameters are all complex numbers.

In contrast, under the irradiation of LPL, the second term in the effective Hamiltonian Eq.~(\ref{Eq.4}) becomes zero, i.e. $\Delta H=0$. This means that LPL will only renormalize the original hopping parameters and not introduce additional hoppings.

\section{Phase transitions}\label{sec:III}

In the following, we investigate the topological phase transition of the ML-TiSiCO by applying a periodic field of laser and a gate-electric field. By adjusting the chirality and intensity of the incident light, as well as the magnitude and direction of the electric field, we can obtain various topological phases.

\begin{figure}
\includegraphics[width=1\columnwidth]{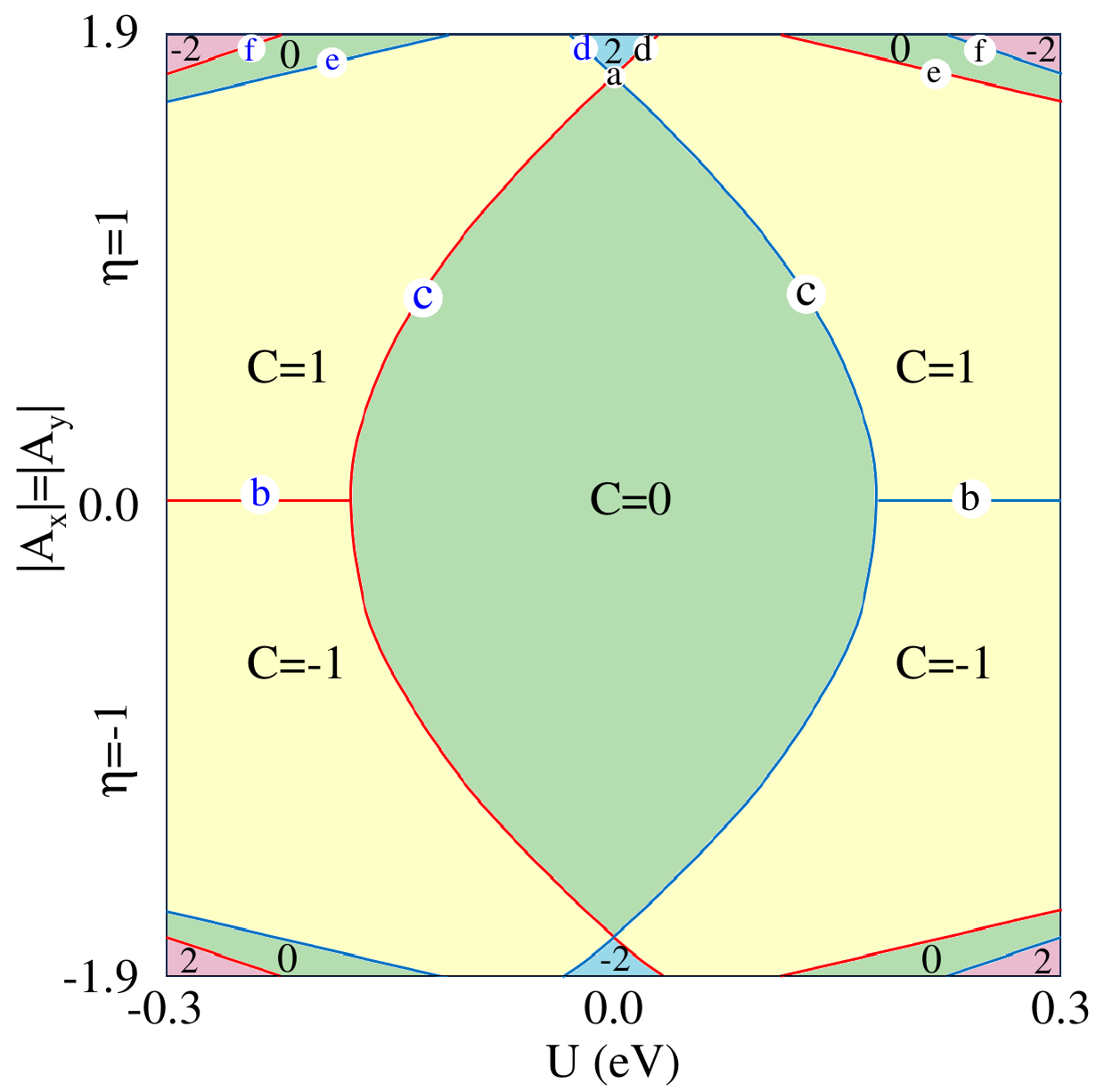}
\caption
{Topological phase diagram of ML-TiSiCO driven by the CPL with high frequency $\omega=3W$ and perpendicular electric field. $\eta=1,-1$ represent the right-handed or left-handed polarized light, respectively. The black and blue letter represent the topological semimetal phase under the positive or negative gate-electric field, respectively.}
\label{Fig.2}
\end{figure}

\subsection{Topological phase transitions under CPL}

\begin{figure*}
\includegraphics[width=1.6\columnwidth]{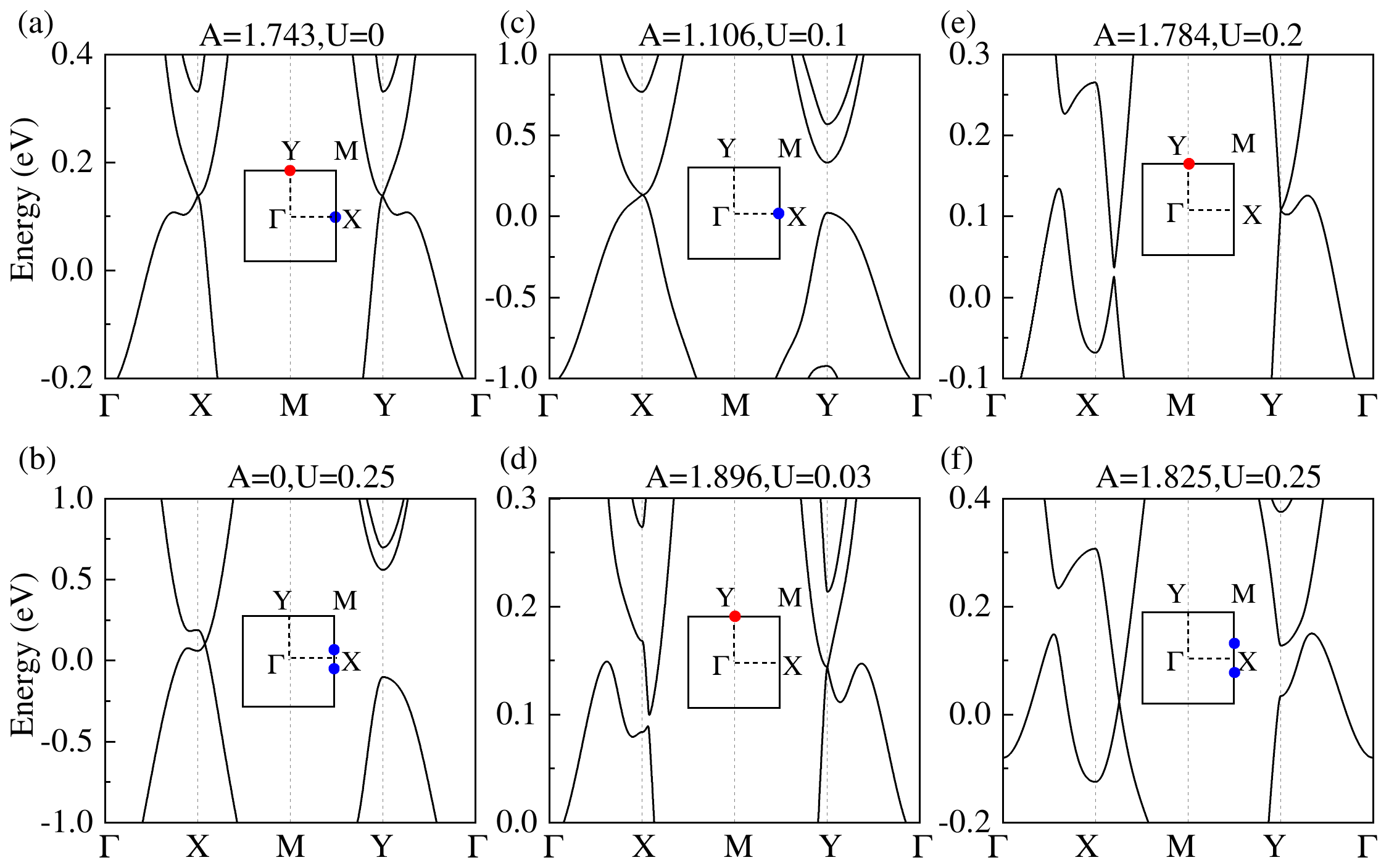}
\caption
{The band structures of the semimetal phases for ML-TiSiCO under CPL and positive electric field ($U>0$) with different intensities, correspond to the black letters marked in Fig.~\ref{Fig.2}. The red and blue dots mark the positions of band crossing points in the BZ.}
\label{Fig.3}
\end{figure*}

\begin{figure*}
\includegraphics[width=1.6\columnwidth]{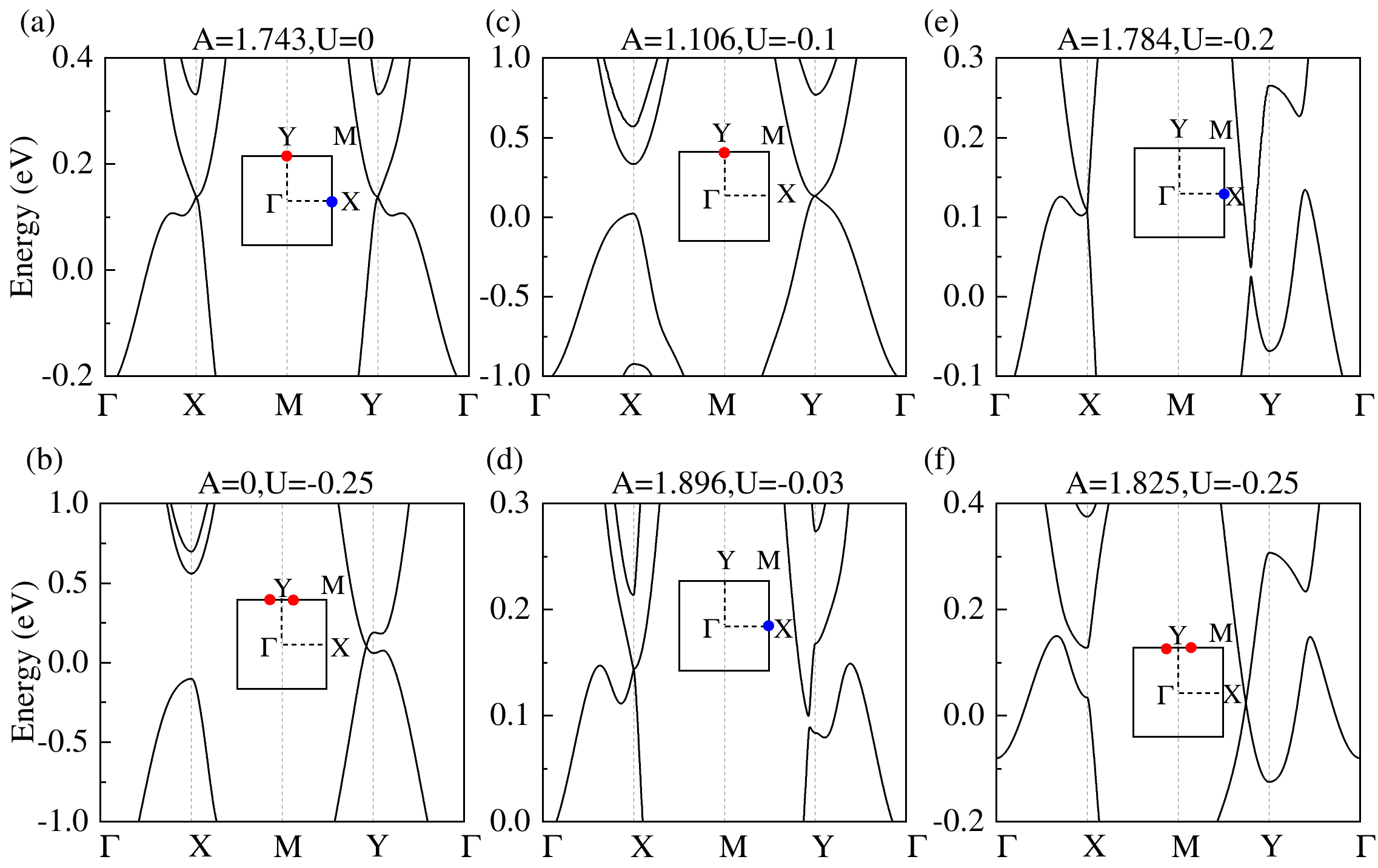}
\caption
{The band structures of the semimetal phases for ML-TiSiCO under CPL and negative electric field ($U<0$) with different intensities, correspond to the blue letters marked in Fig.~\ref{Fig.2}. The red and blue dots mark the positions of band crossing points in the BZ.}
\label{Fig.4}
\end{figure*}

We first investigate the influence of CPL on the ML-TiSiCO. By controlling the intensity $\mathbf{A}(t)$ and chirality $\eta$ of CPL, as well as the magnitude and direction of the electric field $E_z$, we can obtain a rich phase diagram that includes various topological phases, as shown in Fig.~\ref{Fig.2}. In Fig.~\ref{Fig.2}, the green, yellow, and pink regions correspond to topological insulators with Chern numbers of $C=0$, and Chern insulators with Chern numbers of $C=\pm$1 and $C=\pm$2, respectively. During the transition between these topological phases with different Chern numbers, the conduction and valence bands will cross, resulting in different topological semimetals. The blue and red lines in Fig.~\ref{Fig.2} are phase boundaries, which also represent the crossing points near the $X$ and $Y$ points in the BZ, respectively. Additionally, the black and blue letters marked in Fig.~\ref{Fig.2} represent the system under a positive ($U>0$) and negative ($U<0$) perpendicular electric field, corresponding to the band structures in Fig.~\ref{Fig.3} and~\ref{Fig.4}, respectively.

We first focus on the topological phase transitions of ML-TiSiCO with zero electric field ($U=0$), while adjusting the intensity of the right-handed CPL ($\eta=1$). Because the CPL incident along the $z$ direction does not break the $S_{4z}$ symmetry of the system, the $X$ and $Y$ valleys always remain degenerate. As the intensity of CPL increases, the band gaps of both the $X$ and $Y$ valley simultaneously decrease until the conduction and valence bands linearly cross, forming two Dirac points. The two Dirac points are located at $X$ and $Y$ points, as shown in the band structure in Fig.~\ref{Fig.3}(a), where the positions of the two Dirac points are marked in the BZ. This critical point is labeled with the black letter ``a'' in Fig.~\ref{Fig.2}. With increasing the intensity of CPL, the band gaps at both the $X$ and $Y$ valleys will reopen, and during this process, band inversion occurs in both valleys. Then, the system undergoes a topological transition from a second-order topological insulator to a Chern insulator, with the Chern number changing from $C=0$ to $C=2$.

We fix the intensity of CPL as $A=0$ and only adjust the magnitude of the electric field. The perpendicular electric field breaks the $S_{4z}$ symmetry, resulting in the $X$ and $Y$ valleys no longer being equivalent. As the electric field magnitude increases ($U>0$), the band gap at the $X$ valley gradually decreases, while the band gap at the $Y$ valley increases~\cite{ChaoxiCui}. When the electric field continues to increase to a critical value $U=U_c$, the valence and conduction band will touch at $X$ valley and form a semi-Dirac point. When the electric field $U>U_c$, the semi-Dirac point splits into two Dirac points located at the $M$-$X$ path and both Dirac points reside around the $X$ valley [see the BZ in the Fig.~\ref{Fig.3}(b)]. Therefore, we can obtain valley polarized topological Dirac semimetals, which are shown in Fig.~\ref{Fig.3}(b), corresponding to the black ``b'' labeled in Fig.~\ref{Fig.2}.

Next, with the electric field fixed at $U<U_c$, we adjust the intensity of the right-handed CPL($\eta=1$). We already know that when the electric field $U<U_c$, the system still exhibits a global band gap, but the $X$ valley and $Y$ valley are no longer being equivalent. The CPL will decrease the band gaps of both the $X$ and $Y$ valleys, but it will first make the conduction and valence bands touch at the $X$ valley, forming a Dirac point. The critical point is labeled with the black letter ``c'' in Fig.~\ref{Fig.2}, and the corresponding band structure is shown in Fig.~\ref{Fig.3}(c). As the intensity of CPL continues to increase, the band gap of $X$ valley reopens, accompanied by a band inversion, while the band gap in $Y$ valleys remains greater than 0. The Chern number changes from $C=0$ to $C=1$. With the increase in the intensity of CPL, there will also be a band inversion occurring in the $Y$ valley, leading to changes in the Chern number. During the band inversion process, the conduction and valence bands will also touch at the $Y$ valley, forming a Dirac point. However, depending on the magnitude of the electric field, there will be two distinct topological phase transitions. When the electric field is very small and close to 0, the band inversion in the $Y$ valley will make the Chern number change from $C=1$ to $C=2$. This critical point is labeled with black ``d'' in Fig.~\ref{Fig.2}, and its band structure is illustrated in Fig.~\ref{Fig.3}(d).
Just like the band structure in Fig.~\ref{Fig.3}(e), when the electric field is relatively strong, although the change of band structure is similar to that of the small electric field, the Chern number changes from $C=1$ to $C=0$.

Subsequently, with the electric field fixed at $U>U_c$, we continue to adjust the intensity of right-handed CPL($\eta=1$).
When the electric field $U>U_c$, the band gap at $X$ valley has already closed and the valley polarized Dirac points have formed. Therefore, once the system is irradiated by CPL, the band gap of the $X$ valley will be directly opened, and the Chern number become $C=1$.  As the intensity of CPL continues to increase, the band gap of the $Y$ valley gradually decreases to zero, forming a crossing point which is shown in the Fig.~\ref{Fig.3}(e). When the band gap of the $Y$ valley opens, the system's Chern number changes from $C=1$ to $C=0$.
As the intensity of CPL increases, while the band gap of $Y$ valley still opens, the conduction and valence bands will once again linearly cross at the $M$-$X$ path and form two Dirac points. The band structure are shown in the in Fig.~\ref{Fig.3}(f) and the position of the crossing points are marked in the BZ. Both Dirac points reside around the $X$ valley, making this phase a valley-polarized Dirac semimetal. 
Further increasing $A$, the CPL opens the band gap of the two Dirac points and the system's Chern number changes from $C=0$ to $C=-2$.

If the chirality of incident light is reversed ($\eta=-1$), under the left-handed CPL and perpendicular electric field, the band structure will still undergo the aforementioned changes, but the sign of the Chern number will be reversed. If the direction of the electric field is reversed ($U<0$), the changes in the Chern number will be the same as the system under the positive electric field $U>0$, but the band variations at the $X$ and $Y$ valleys exhibit an opposite trend compared to the case of $U>0$.
The band structures of topological semimetal phases generated by the interaction of right-handed CPL ($\eta=1$) and a negative electric field ($U<0$) are shown in Fig.~\ref{Fig.4}, corresponding to the blue letter labeled in Fig.~\ref{Fig.2}.

\begin{figure*}
\includegraphics[width=1.6\columnwidth]{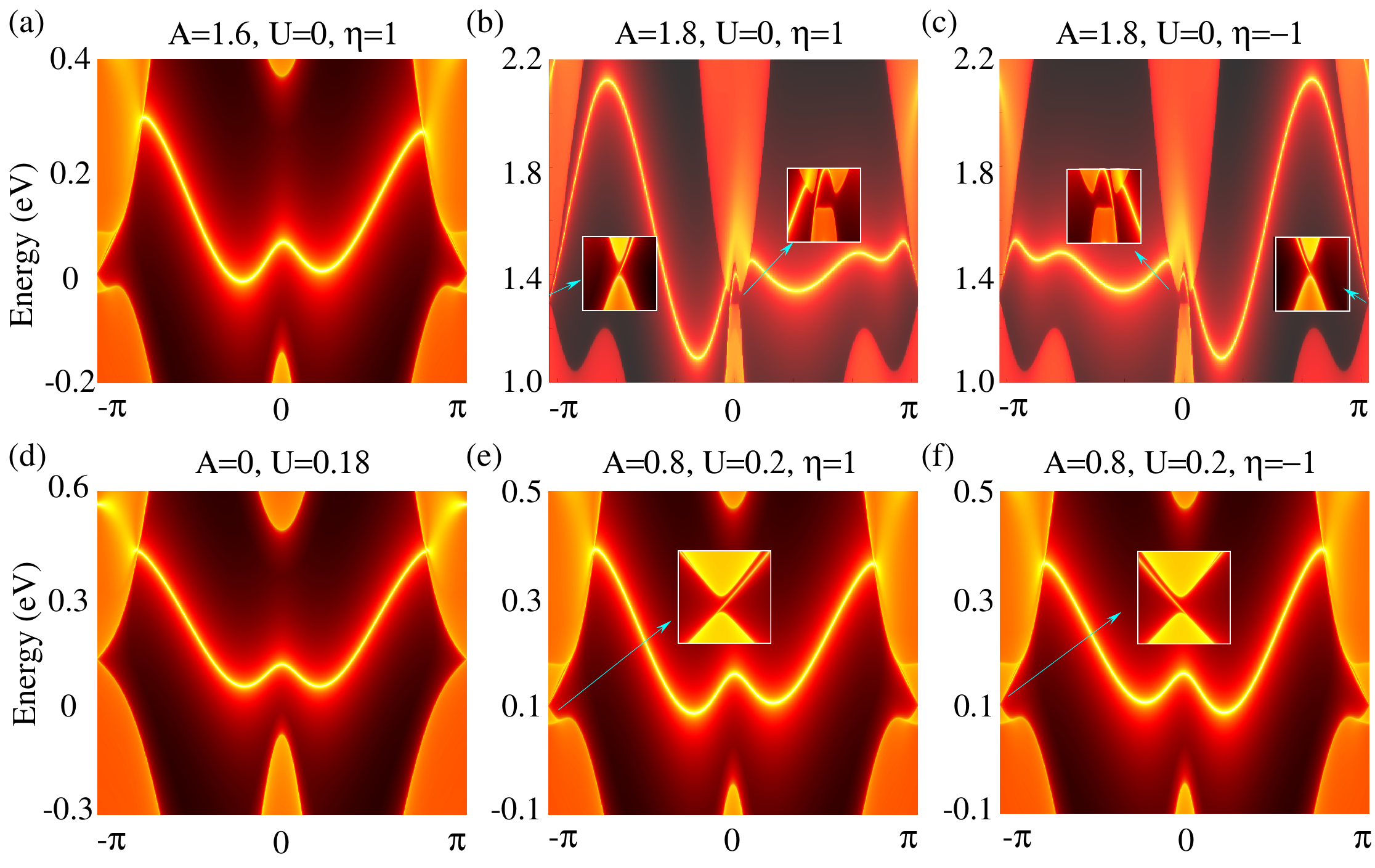}
\caption
{The band structure of the ML-TiSiCO nanoribbon extended in [$100$] direction driven by CPL and gate electric field. The details of the edge states are shown in the inset of the figures.}
\label{Fig.5}
\end{figure*}

The bulk-edge correspondence allows for a deep understanding of the changes in the Chern number, which is associated with topological phase transitions. Under CPL and perpendicular electric field, the ML-TiSiCO changes from a second-order topological insulator to the Chern insulator with different Chern numbers. Here, we investigate the band structure of the nanoribbon of the ML-TiSiCO extended in the [100] direction. Under the irradiation of CPL, as the intensity of CPL increases, the band gap decreases.
From Fig.~\ref{Fig.5}(a), we found before the band inversion, there will be an edge state spanning the entire first BZ, which is protected by $\pi$ Zak phase~\cite{YilinHan}.

As the intensity of CPL increases, band inversion will occur at both the $X$ and $Y$ valleys, resulting in a Chern number of $C=2$. Consequently, two chiral edge states can be observed in Fig.~\ref{Fig.5}(b), where the insets show the details of the edge states. If we change the chirality of the incident laser, i.e., switch to left-handed CPL, with the same intensity of light, two non-trivial edge states will still appear, which is shown in Fig.~\ref{Fig.5}(c). However, their chirality will be opposite to that of the right-handed CPL irradiated system. This confirms that the Chern number of the driven system is switched by the chirality of the incident light.

The positive electric field ($U>U_c$) will make the conduction and valence bands touch at $X$ valley, but it increases the band gap in $Y$ valley, which can also be observed in Fig.~\ref{Fig.5}(d).
When the direction of the electric field is reversed ($U<0$), the band structure changes in the $X$ valley and $Y$ valley also exhibit opposite behaviors to the case of $U>0$.
From the previous discussion, we known that, after the band gap is closed by the electric field $U>U_c$, the irradiation of right-handed CPL ($\eta=1$) will open the band gap and the Chern number becomes $C=1$. A chiral edge state appears in the band of the nanoribbon extended in [100] direction as well. If we switch the incident light to left-handed CPL ($\eta=-1$), the chirality of the edge states will become opposite, corresponding to a reversal Chern number $C=-1$. For the both case of $\eta=\pm 1$, the chiral edge state is shown in Fig.~\ref{Fig.5}(e-f).

\subsection{Topological transitions under LPL}

\begin{figure*}
\includegraphics[width=1.6\columnwidth]{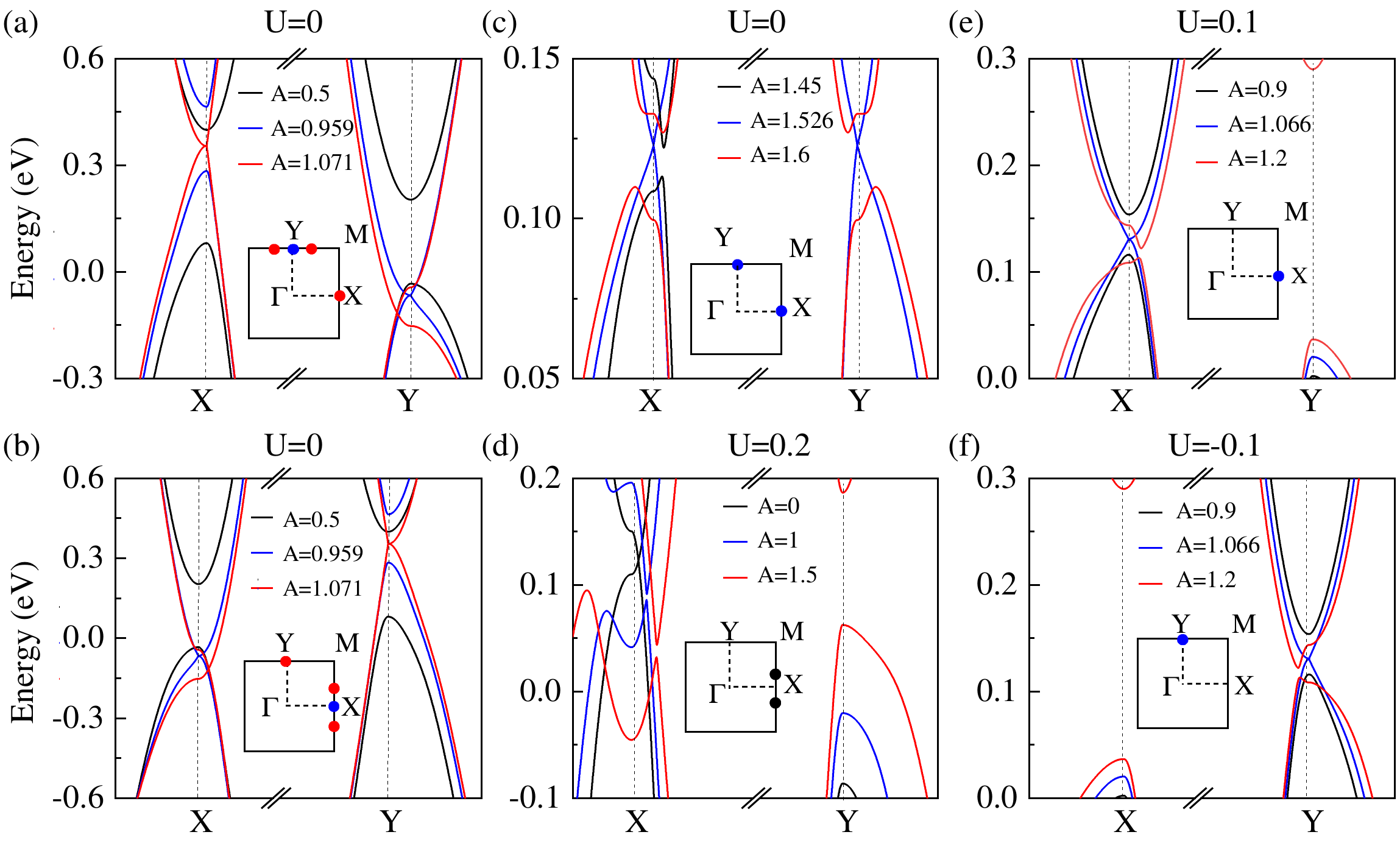}
\caption
{The band structures of ML-TiSiCO under LPL and gate electric field. The LPL is incident along the (a) $x$, (b) $y$, and (c-f) $z$ direction. The black, red, and blue dots in the BZ represent the positions of the band crossing points, and they correspond to the band structures with the same color.}
\label{Fig.6}
\end{figure*}

In this subsection, we study the effect of LPL on the energy bands of ML-TiSiCO. The LPL does not break the time-reversal symmetry of ML-TiSiCO, and thus, it does not change the Chern number of the system. However, LPL still significantly modulates the band structure and leads to topological phase transitions.
We consider the impact of LPL propagating along the $x$, $y$, and $z$ three directions on the band structure of ML-TiSiCO. The band structures are illustrated in Fig.~\ref{Fig.6}. Here, the black, blue, and red dots indicate the positions of band crossing points in the BZ, and they are associated with the bands plotted by the same color.

When the incident light propagates along the $x$ or $y$ direction, the $S_{4z}$ symmetry of the system is broken, resulting in the non-equivalence of the $X$ and $Y$ valleys. Consequently, the response of both valleys to an external field is no longer the same. When the incident light propagates in the $x$ direction, with increasing the intensity of LPL, the energy of the $X$ valley rises, while the energy of the $Y$ valley decreases, and the local band gaps of both valleys reduce. The above changes in the band structure are shown in Fig.~\ref{Fig.6}(a). With the increase in the intensity of LPL, the bandgap of the $Y$ valley first closes, forming a semi-Dirac point, which is plotted by the blue line in Fig.~\ref{Fig.6}(a), and the positions of the crossing points are also marked with blue dots in the BZ. Subsequently, the semi-Dirac point splits into two Dirac points located at $M$-$Y$ path, which is valley polarized Dirac semimetal phase. As depicted by the red line in Fig.~\ref{Fig.6}(a), with the continued increase in the intensity of LPL, the band gap at $X$ valley will also closes and a semi-Dirac point forms at $X$ point, while there are still two Dirac points around $Y$ valley. The locations of these crossing points are also indicated by red dots in the BZ. The semi-Dirac point at the $X$ valley will further split into two Dirac points located at $M$-$X$ path, and the four Dirac points near the two valleys will no longer be opened by LPL.
That is, the LPL incident along the $x$ direction can induce a phase transition from a topological insulator to a topological semimetal. Furthermore, the position and type of the crossing points can be controlled by the intensity of the incident LPL.
When LPL is incident along the $y$ direction, as the intensity of LPL increases, the energy of the $X$ valley decreases while the energy of the $Y$ valley rises. Additionally, the other changes in the bands of the $X$ and $Y$ valleys are opposite to the system driven by the LPL along the $x$ direction.
Eventually, two Dirac points will also form at both the $M$-$X$ and $M$-$Y$ paths. This process is illustrated in Fig.~\ref{Fig.6}(b).

When the LPL is incident along the $z$ direction, the $S_{4z}$ symmetry of the system does not be broken, and the $X$ and $Y$ valleys remain degenerate. Therefore, with an increase in the intensity of LPL, the band gaps of both the $X$ and $Y$ valleys simultaneously decrease to zero, forming two Dirac points. In Fig.~\ref{Fig.6}(c), we found as the intensity of LPL increases, the band gaps of the two Dirac points will open. During this process, the Chern number remains unchanged, but the system transitions from a topological insulator to a Dirac semimetal.

When the LPL and electric field are both along the $z$ direction, the $X$ and $Y$ valleys are no longer equivalent. When the electric field $U>U_c$, the band gap at $X$ valley closes and the valley polarized Dirac semimetal phase form, which is plotted by the black line in Fig.~\ref{Fig.6}(d).
The Dirac points also can be opened by LPL and the corresponding band structures are plotted by the blue and red lines in Fig.~\ref{Fig.6}(d).

When the electric field $0<U<U_c$, with the increase in the intensity of LPL, the band gap of the $X$ valley will first decrease to zero. In Fig.~\ref{Fig.6}(e), the conduction and valence bands touch at $X$ point and form a Dirac point. The Dirac point will be opened by the LPL with larger intensities.

Conversely, if the electric field direction is reversed ($U<0$), there is a Dirac point formed in $Y$ valley, as shown in Fig.~\ref{Fig.6}(f). Therefore, under the control of LPL and electric field, even though the Chern number of the system remains unchanged, it can transition from a topological insulator to a Dirac or simi-Dirac semimetal. Further, the positions and types of the crossing points can be controlled by the direction and intensity of LPL and electric field.

\section{Conclusions}\label{sec:IV}

In conclusion, based on the four-band TB Hamiltonian and Floquet theory, we have systematically studied the topological phase transitions of ML-TiSiCO under the periodic driving polarized light and gate-electric field. Via the effective Floquet Hamiltonian, we found the influence of CPL on the system is equivalent to renormalizing the original hoppings parameters and introducing more neighboring hoppings. But LPL only modulates the origin hopping parameters.
Under the control of CPL and perpendicular electric field, we obtained a rich phase diagram that includes Chern insulators with different Chern numbers and various topological semimetals. The Chern number of the Chern insulators or the position and type of the crossing points of a topological semimetal can both be adjusted by the intensity of CPL and electric field. Furthermore, we also observed chiral edge states in the band structure of the nanoribbon extended in the [100] direction. The chirality of the edge states and the Chern number of the driven system can be switched by the chirality of the CPL.

The LPL does not break the time-reversal symmetry, which means that under the driving of LPL, the Chern number of the system remains unchanged. However, the system can still transition from a second-order topological insulator to various topological semimetals, including semi-Dirac, Dirac, and valley polarized Dirac semimetals. The type of topological semimetal and the position of its crossing points in BZ can be manipulated by both adjusting the intensity and direction of LPL and the electric field. 
%Our work not only deepens the understanding of the TiSICO family of materials but also broadens the potential applications of these materials in non-equilibrium optoelectronic devices.

\begin{acknowledgments}
This work was supported by the NSF of China (Grants No. 12004035, No. 12234003, and No. 12061131002).

\end{acknowledgments}

\bibliography{photo_TiSiCO_v1120}
\end{document}